\documentclass[11pt,twoside]{book} 
\usepackage{asp2008s}
\usepackage{times}
\usepackage{lscape}
\usepackage{epsf}

\usepackage{graphicx}
\usepackage{amssymb}
\usepackage{longtable}
\usepackage[figuresright]{rotating}
\usepackage{multirow}

\newcommand{\simless}{\mathbin{\lower 3pt\hbox {$\rlap{\raise 5pt\hbox{$\char'074$}}\mathchar"7218$}}}

\mainmatter

\newlength{\deftabcolsep}
\begin{document}

\title{The Embedded Massive Star Forming Region RCW 38}   

\author{Scott J. Wolk}   
\affil{Harvard--Smithsonian Center for Astrophysics, Cambridge MA 02138,
USA}    

\author{Tyler L. Bourke}   
\affil{Harvard--Smithsonian Center for Astrophysics, Cambridge MA 02138,
USA}    

\author{Miquela Vigil}   
\affil{Lincoln Laboratory, Massachusetts Institute of Technology, Lexington
MA 02420, USA}    

\begin{abstract} 
RCW~38 is a uniquely young ($<$1 Myr), embedded ($A_V \sim 10$) stellar
cluster surrounding a pair of early O stars ($\sim$O5.5) and is one of the
few regions within 2~kpc other than Orion to contain over 1000 members.
X-ray and deep near-infrared observations reveal a dense cluster with over
200 X-ray sources and 400 infrared sources embedded in a diffuse hot plasma
within a 1~pc diameter.  The central O star has evacuated its immediate
surroundings of dust, creating a wind bubble $\sim$0.1 pc in radius that
is confined by the surrounding molecular cloud, as traced by millimeter
continuum and molecular line emission. The interface between the bubble and
cloud is a region of warm dust and ionized gas, which shows evidence for
ongoing star formation.  Extended warm dust is found throughout a  2--3 pc
region and coincides with extended X-ray plasma.  This is evidence that the
influence of the massive stars reaches beyond the confines of the O star
bubble. RCW~38 appears similar in structure to RCW~49 and M~20 but is at an
earlier evolutionary phase.  RCW~38 appears to be a blister compact
H{\small II} region lying just inside the edge of a giant molecular cloud.

\end{abstract}


\begin{figure}[h]
\begin{center}
\includegraphics[draft=False,scale=0.2,angle=0]{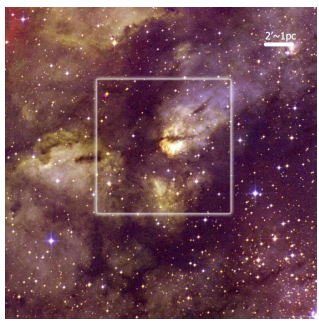}
\end{center}
\caption{An optical image of RCW 38, about 20\arcmin\ ($\sim$10 pc) on a
side, based on images from the digitized sky survey (DSS).  Blue plates are
printed as blue, red plates are printed as yellow, near-infrared data are
printed in red.  The region in the green box is shown at mid-infrared
wavelengths in Figure~2.}
\label{dss}
\end{figure}

\begin{figure}[h]
\begin{center}
\includegraphics[draft=False,scale=0.50,angle=0]{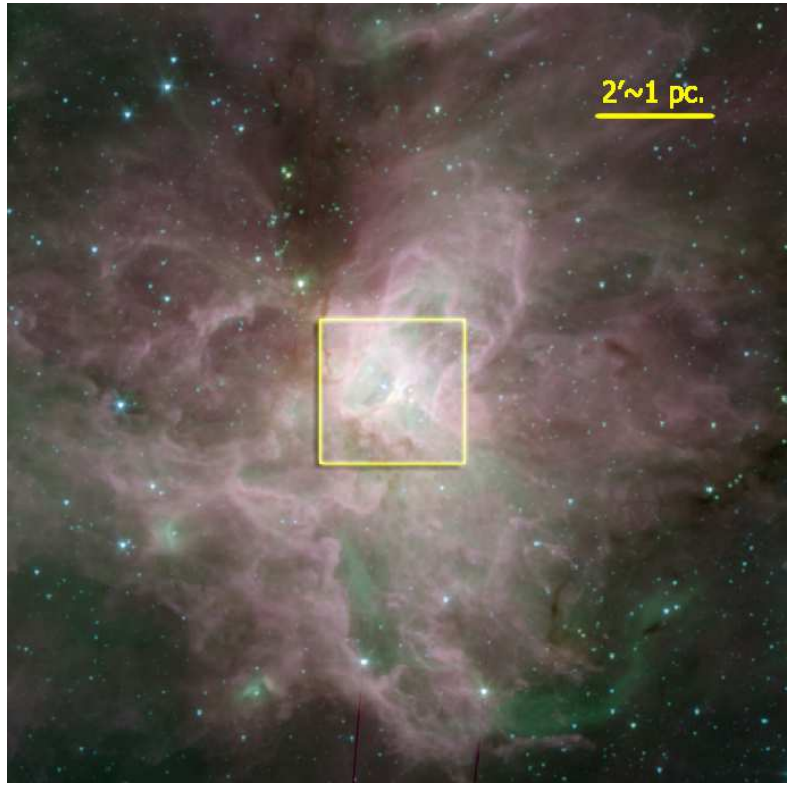}
\end{center}
\caption{Mid-infrared image of RCW 38, about 11.5\arcmin\ (5.7 pc) on a
side, from $Spitzer$/IRAC observations.  IRAC Band 1 (3.6 \micron) is
assigned to blue, Band 2 (4.5 \micron) to green, and Band 3 (5.6 \micron)
to red.  [3.6] band = blue, [4.5] band = green and [5.6] band = red.  Voids
in the 5.6 \micron\ image tend to be filled by 4.5 \micron\ emission.  The
yellow box is 2.5\arcmin\ on a side and is shown in more detail in
Figures~3 and 5.}
\label{2MASS}
\end{figure}

\section{Introduction}   

The evolution of high mass clustered star forming regions is complex and
poorly understood.  Only the nearby ($\sim$400 pc), optically revealed, Orion
Nebula Cluster (ONC) is well studied (see Muench et al.\ and  O'Dell et al.\
in this Handbook).  Yet, a wide variety of high mass embedded clusters is found within
2 kpc of the Sun.  Within this limit, the young cluster RCW~38
(08$^h$59$^m$47.2$^s$ -47$^\circ$31$'$57$''$ (J2000), {\it l,b} =
$268.03^\circ,-0.98^\circ$) is one of the few regions other than the ONC to
contain over 1000 members (Lada \& Lada 2003; Wolk et al.\ 2006).  RCW~38
has an embedded and dense stellar population, comparable to other regions
that have been studied recently with Spitzer (e.g., M~20 and RCW~49; Rho et
al.\ 2004, Whitney et al.\ 2004, Churchwell et al.\ 2004).  RCW~38 provides
a unique opportunity to study the evolution of a rich cluster during the
phase where its most massive members, a pair of O5.5 stars (DeRose et al.\
2008) have just completed their ultracompact {H}{\small II}\ region
(UC{H}{\small II}) phase and are now greatly influencing its natal
environment and the evolution of its low mass members.

\section{Overview}

\begin{figure}[!ht]
\begin{center}
\includegraphics[draft=False,scale = 0.5, angle = 0]{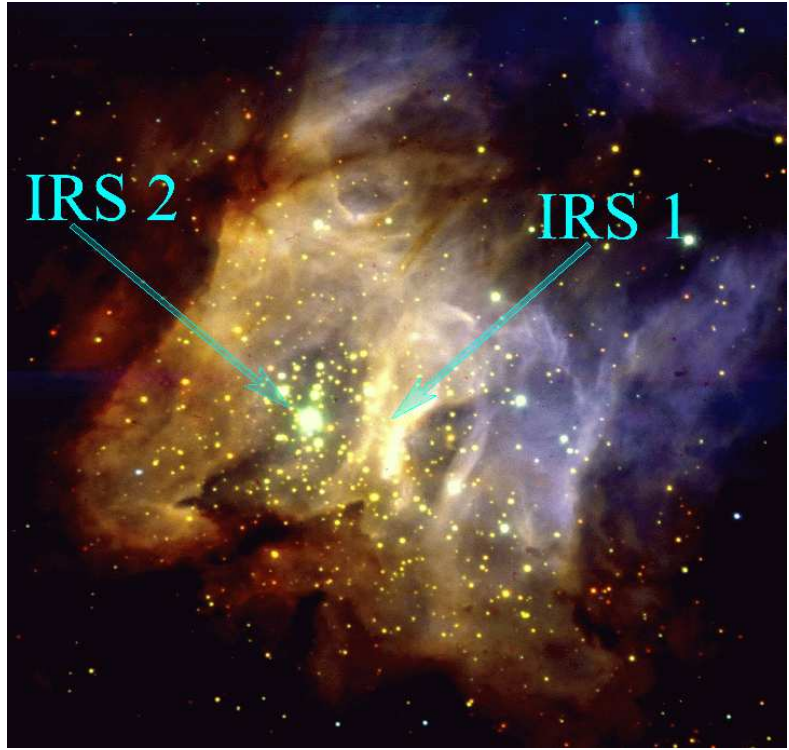}
\end{center}
\caption{A close-up of the central 2.5\arcmin\ ($\sim$1.2 pc) of RCW~38
(Wolk et al.\ 2006). In this VLT image, Z band data are printed as blue, H
band data are green and K band are red.  IRS~2 is the brightest point
source in the field.  IRS~1 is associated with the bright ridge of emission
to the west of IRS~2, near the center of the image.  The diffuse radiation
is a mixture of starlight scattered by the dust and gas in the area, and
atomic and molecular hydrogen line emission. Essentially every star visible
in this image is a cluster member (Wolk et al.\ 2006; DeRose et al.\
2008).}
\label{VLT}
\end{figure}

RCW 38 was catalogued in both the H$\alpha$ survey by Rodgers, Campbell \&
Whiteoak (1960) and the earlier survey of southern {H}{\small II}\ regions
(Gum 1955) as a moderately bright region of emission about 40\arcmin\ on a
side (Figure~\ref{dss}). Radio surveys during the 1960's indicated that it
was one of the brightest {H}{\small II}\ regions at radio wavelengths
(e.g., Wilson et al.~1970).  The radio brightness made it an early
candidate supernova remnant until its spectrum was shown to be thermal.  A
complicating factor in the study of RCW~38 is that it is adjacent to the
Vela Molecular Ridge (see the chapter by Pettersson).  In fact, there is a
faint ring shape structure -- perhaps a supernova remnant (SNR) -- in
apparent contact with the star forming cloud (RX J0852.0-4622).  However,
this seems to be a chance superposition as the SNR is about a factor of two
closer than RCW~38 based on absorption arguments (Aschenbach et al.\ 1999).
The 5 GHz survey by Shaver \& Goss (1969) shows RCW~38 as both a bright
continuum source at multiple wavelengths (Shaver \& Goss 1970) and broad in
its extent of greater than 30\arcmin\ in diameter.  It was not until the
early 1970's that this region was clearly associated with massive young
stars (Johnson 1973).  The next year, with improving resolution and
detectors, H90$\alpha$ observations indicated that the radio peak of the
region had a ring--like shape about 2\arcmin\ across (Huchtmeier 1974).
Figure~\ref{dss} shows an optical view of the RCW~38 region.  Note the
central region is opaque due to high extinction.  This has confined the
study of the central region mostly to infrared and radio wavelengths.
Several infrared studies indicated a ``ring--like'' or ``horseshoe''
structure around IRS~2 (Figures~\ref{2MASS},\ref{VLT}), which is also seen in
high resolution radio and millimeter wavelength studies (Vigil 2004; Wolk
et al.\ 2006).  These structures are shown in detail in Figure~\ref{VLT}.
There are two defining infrared sources in RCW~38 (Frogel \& Persson 1974)
-- the brightest at 2 \micron\ is labeled IRS~2 and the brightest at 10
\micron\ is referred to as IRS~1.

\section {IRS 1}

Frogel \& Persson (1974) studied RCW~38 at 10 $\mu$m with moderate spatial
resolution (14.5\arcsec) and found that the warm dust emission follows a
horse-shoe shape across a $1.8 \times 1.8$ arcmin$^{2}$ area; they labeled
the brightest unresolved peak in its western side IRS~1.  Epchtein \& Turon
(1979) mapped the central region of RCW~38, at 10~\micron, with about twice
the resolution of Frogel \& Persson and resolved IRS~1 into several
discrete sources; they suggested that this may be a cluster of
embedded stars, younger than those associated with IRS~2.

In MSX observations at 8.3 $\mu$m the central region of RCW~38
exhibits a circular emission structure, perhaps slightly elongated about
2\arcmin\ across, in agreement with the earlier mid-infrared studies.  This
is resolved by $Spitzer$/IRAC observations at 3.6-8.0 $\mu$m (Wolk PI) to
show a roughly heart shaped region of polycyclic aromatic hydrocarbon (PAH)
emission which includes numerous bubbles filled with 4.5~\micron\ diffuse
emission (see Figure~\ref{2MASS}).  The central region appears to lie
within a large bubble about 0.4~pc across.  The O5.5 binary IRS~2 appears
to lie in the center of the bubble as seen in the mid- and near-infrared
data (Sect.~4 and Figures~2 and~3). The inner face of the eastern
part of the central bubble appears to be illuminated, giving the impression
that we are not looking straight down the opening but rather offset by
between 30 and 45 degrees.

Smith et al.\ (1999) made high resolution mid-infrared maps of the core
region of RCW~38.  They concentrated on the IRS~1 region that has the peak
radio and 10 \micron\ emission.  They found that IRS~1 is a dust ridge
extending 0.1-0.2 parsec predominantly in the north--south direction less
than 0.1~pc west of IRS~2. The dust ridge has a color temperature of about
175~K and has several condensations within it.  They identify IRS~1 as the
brightest peak among them, lying interior to all the other peaks about
12\arcsec\ west of IRS~2.

\section{IRS 2}
\label{sec-irs2}

From its near-infrared (NIR) colors and assuming a distance of 1.5 kpc,
Frogel \& Persson (1974) suggested that the brightest 2~\micron\ source,
IRS~2, is an early O-star (specifically O4) having about 12.8 magnitudes of
visual extinction, or a group of later spectral type O stars.  This is
sufficient to account for the observed continuum emission at 5 GHz (Shaver
\& Goss 1970).  Furniss, Jennings, \& Moorwood (1975) combined broad-band
40-350 $\mu$m measurements with the previous work of Frogel \& Persson,
deriving a total luminosity of 7$\times 10^5$~L$_\odot$ and arguing that an
O4 star would supply more than enough luminosity to account for all the
infrared luminosity and emit enough Ly photons to yield twice the observed
radio flux, unless significant absorption by dust is occurring.  They
suggest however that IRS~2 is an O5 star with little or no dust absorption
of the continuum photons in the ionized region.  The higher resolution
imaging of Ligori et~al.\ (1994) confirmed the presence of a cluster of at
least 5 optically invisible young stellar objects embedded in nebular
emission in the central 1.5\arcmin\ near IRS~2.

Frogel \& Persson (1974) produced detailed 1.25 -- 20 \micron\ mapping of
the region.  At 10~\micron\ they confirm a ring-like structure with the
bright 2 \micron\ source -- IRS~2 -- very near the center and a dominating
10 \micron\ source (IRS~1) about 10\arcsec\ to the west. A third 10
\micron\ source about 40\arcsec\ to the east of IRS~2 was also noted.
Persson et al.\ (1976) revisited this region using narrow band photometry
and found no evidence of silicates.  Other narrow line infrared imaging
such as Br-$\gamma$ at 2.16 \micron\ also indicated an open center
morphology centered on IRS~2 with enhanced emission toward IRS~1 (Mizutani
et al.\ 1987).


Mid-infrared imaging and spectroscopy by Smith et~al.\ (1999) provide
further evidence that the ionizing source of the region is at least O5 or
earlier.  Through modeling of the ratio of emission lines at IRS1, the
ionizing radiation field constrains the ionizing source to be between O4.5
and O5.5.  Smith et~al.\ find that the region surrounding IRS2 is depleted
of dust, suggesting the region can be explained in terms of a wind-blown
cavity, where winds from IRS2 have blown out the surrounding material.

DeRose et al.\ (2008) used adaptive optics imaging to directly show that
IRS2 is a binary with equal mass members and a projected separation of
$\sim$ 500~AU.  The presence of two O stars of equal mass combined with the
earlier data on luminosity and ionization imply that they have spectral
types of O5.5.

\section{Massive Stars}

Muzzio (1979) and Muzzio \& Celotti de Frecha (1979) identified OB stars
throughout the southern sky using photometry and objective prism
photography.  They identified about 20 OB star candidates in the vicinity
of RCW~38. All have absolute V magnitudes less than zero but are
concentrated about 10--20\arcmin\ southwest of IRS~2.   Wolk et al.\ (2006)
use the absolute K-band magnitudes of X--ray sources to determine the
existence of 31 candidate OB stars over a 5 pc$^2$ region centered on
IRS~2.  More than 20 of these are found to be in the central $\sim$1 pc. No
unusual clustering to the southeast was noted.  To make mass estimates they
estimated the age of the cluster at 0.5~Myr and noted that if they assumed
an age of 1~Myr, the number of OB stars would have more than doubled to a
somewhat less reasonable value. However, the Wolk et al.\ sample was X-ray
based and hence biased against finding soft OB stars behind more than 20
A$_V$ of extinction\footnote{Recent X-ray observations indicate that O
stars are capable of producing very hard X-rays (e.g. Gagn\'e et al.\
2005). However, most B stars and about half of the observed O stars are
still soft sources and hence observations in dusty regions are biased
against detecting them.}.  The OB candidates remain to be confirmed.

\section{Extinction}

NIR imaging and photometry of the region by Storey \& Bailey (1982) showed
a number of highly reddened point sources and they deduced an A$_V$ of
order 60 to the cluster. The 2MASS data show that the nebulosity associated
with RCW 38 is extensive across a large area, with dust lanes and patches
running throughout. Wolk et al.\ (2006) measure the extinction and hydrogen
column to over a score of sources with good X-ray and NIR data and find
A$_V$ ranging from 3 to 20 with strong evidence of NIR sources embedded in
more than 30 magnitudes of visual extinction. Smith et al.\ (1999) find
that the region around IRS~2 has a gas to dust ratio much lower than 100 to
1.  Thus, this area is a true cavity and not just an extinction effect.
Millimeter radio observations indicate that several hundred magnitudes of
extinguishing material may lie behind the observed stars (see Section~11).
Combining these results with the near--IR extinction observations suggests
that RCW~38 is an embedded blister HII region lying just inside the front
edge of a giant molecular cloud. Further, it appears that the HII region is
compacted by the overlying material and beginning to break out in some
locations.

\section{The Embedded Stellar Cluster}

Even in the remarkable $Spitzer$/IRAC image (Figure~\ref{2MASS}) the scope
and extent of the cluster is hard to ascertain. Only a few dozen sources
are identifiable within the bright nebular emission. Outside the IR nebula
cluster members are difficult to distinguish from background stars.  We
have found that 116 sources in the 2MASS catalog, corresponding to the
region shown in Figure~3, have IR-excesses consistent with an optically
thick disk at K--band.  A number of the cluster members are directly
exposed to IRS~2 in the cavity, and could be undergoing the same fate as
the proplyds in the ONC (DeRose et al.\ 2008).  Further, about 250 sources
in the regions are Class~I or Class~II objects based on their IRAC colors.
This group includes the 2MASS selected candidates.  Most of these are
clearly associated with RCW 38. However, at least 2 are associated with a
background star forming region -- BRAN 231A. Coincidences among the X-ray
detected sources and NIR excess sources are found over 10\arcmin\ from
IRS~2 indicating that the cluster extends at least 5 pc.

The first hint of the extent of the low mass star population is provided by
the spectacular NIR image of the heart of the region (Figure~\ref{VLT}).
Despite the small (2.5\arcmin\ on a side) field of view, one can
immediately ascertain several important features of the system from this
image.  First, the blue to red gradient indicates a steep dust extinction
gradient increasing from northwest to southeast.  Second, two regions are
cleared of dust, one about 0.1~pc in diameter centered on IRS~2, and
another of similar size, just west of IRS~1.  Both of these cleared regions
are visible in the optical plates indicating that extinction is not
particularly high for these regions. The bright ridge separating the two
cleared regions is the IRS~1 ridge which was mapped in detail by Smith et
al.\ (1999).  Third, there is a plethora of stars in this image.  Over 480
stars are visible in this image and the vast majority of these are likely
to be cluster members (Wolk et al.\ 2006, DeRose et al.\ 2008).  The data
are complete to the brown dwarf limit for extinctions less than an A$_V$ of
20 (Wolk et al.\ 2006).  About 130 of the NIR sources have X--ray
counterparts -- there are fewer than a dozen X-ray sources without NIR
counterparts in the central 2.5\arcmin.  This leaves about 350 NIR detected
cluster members not detected in X-rays -- an incompleteness of over 70\%.
The X-ray counterparts demonstrate both membership in the cluster and
masses of more than about 0.5 M$_\odot$.  Overall about 345 X-ray sources
detected by $Chandra$ in a 17\arcmin x 17\arcmin\ field centered here are
associated with RCW~38 (Wolk et al.\ 2006).  Completeness arguments based
on the cluster distance, obscuration, and X-ray sensitivity lead to a total
estimated cluster size of between 1500 and 2400 stars.  This makes RCW~38
the largest embedded cluster by membership within 2 kpc of the Sun, after
the ONC.

\section{Diffuse X-ray Emission}

\begin{figure}[t]
\begin{center}
\includegraphics[draft=False,scale=0.6, angle=0]{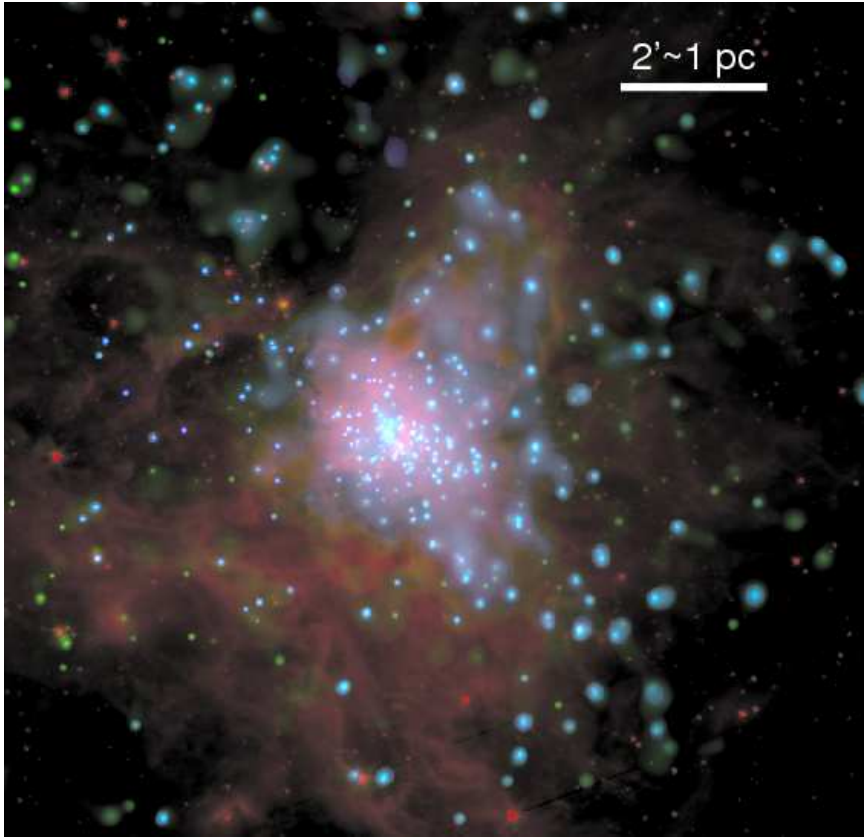}
\caption{A combined X-ray and IR view of the inner 11.5\arcmin\ (5.7~pc.) of RCW 38.
The X-ray data have been adaptively smoothed to about 10\arcsec\
resolution.  Soft X-rays (0.5 -- 1.5 keV) are shown in green and harder X-rays
(1.5 -- 8 keV) in blue. The various structures in the diffuse
X-ray emission seem to trace gaps in the polycyclic aromatic hydrocarbon
(PAH) structures apparent in the 5.8~\micron\ $Spitzer/IRAC$ data (red).
The lack of X-ray emission to the southeast is likely due to high extinction.}
\label{diffuse}
\end{center}
\end{figure}

As shown in Figure~\ref{diffuse}, RCW~38 hosts very bright diffuse X-ray
emission (Wolk et al.~2002).  This X-ray emission differs from that of most
other massive star forming regions because of its power-law spectrum
indicative of synchrotron, not thermal, emission.  Since the spectral
signature is not that of stellar coronae this emission cannot be the sum of
unresolved stars.  Further, it implies the existence of a strong magnetic
field.  Synchrotron emission requires an electron population to be driven
along a magnetic field.  Indeed a significant magnetic field has been
measured in the gas associated with RCW~38 (38$\pm$3 $\mu$G; Bourke et al.\
2001).  Astrophysically, this is most common along the expanding shock
front of a supernova remnant or during violent accretion onto a compact
object with a large magnetic field.  We noted earlier the existence of a
possible supernova remnant in the field, it is offset from RCW~38 by about
6\arcmin\ and there is known X-ray emission associated with this location.
Further, the SNR is foreground to RCW~38 and the diffuse emission becomes
more absorbed towards the southeast.  This is the same pattern seen among
the stars and nebulosity associated with RCW~38 -- indicating that the
diffuse X-ray emission is co--located with the stars and dust of RCW~38.

RCW~38 is the first and currently only known source of diffuse synchrotron
X-ray emission from a very young star forming region (Feigelson et al.\ 2007).
The morphology of the diffuse X-ray emission is striking. It is strongest
in the central region near IRS~2 where radio and IR-nebular emission
appears cleared. Further it is confined on the southeast along a ridge
that also confines the mm continuum (cool dust) emission.  There is also a
distinctive kink where the diffuse X-ray emission is deflected around the
dust ridge near IRS~1. It also appears that the plasma is breaking out of
some form of confinement to the northwest where it traces around the
diffuse NIR emission -- akin to the way a stream moves around a stone.
This break out to the northwest is visible as reflection nebulosity in
Figure~1 and corresponding bubble-like structures in Figure~2.

\section{The Distance to RCW 38}

The first reasonable estimate for the distance was from Radhakrishnan
et al.\ (1972).  They estimate a kinematic distance of 1.5 kpc. Muzzio
(1979) identified about 10 OB star candidates in the vicinity of
RCW~38 and derived a distance of about 1.7 kpc by placing the stars on
the main sequence of an HR diagram.  Murphy (1985) examined the CO
structures in the Vela ridge in addition to the photometric data and
derived a distance of 1.6 $\pm 0.8$ kpc.  Avedisova \& Palous (1989)
performed photometric measurements of 218 star forming regions and
also achieve a result of 1.7 kpc.  The X-ray luminosity function is
consistent with the 1.7 kpc distance (Wolk et al.\ 2006).  At this
distance, 1~arcmin corresponds to 0.495 pc.

\section{Radio Observations}

Early low resolution radio observations of continuum emission and
recombination lines indicate typical HII region conditions, i.e., an
electron temperature of $\sim$8000 K and an emission measure of $\sim 10^6$
pc cm$^{-6}$ (e.g., McGee \& Newton 1981, Caswell \& Haynes 1987).  A
relatively high electron density of $\sim10^4$ cm$^{-3}$ supports the view
that the HII region is confined by the surrounding molecular cloud.
Resolved radio studies reveal a clumpy ring-like structure around IRS~2
showing the general morphology seen in the infrared (Huchtmeier 1974; Vigil
2004).  As in the mid-infrared, the brightest radio peak coincides with
IRS~1 (Figure~5).  High resolution observations indicate the ring is mostly
optically thick at centimeter wavelengths, with no evidence for synchrotron
emission (Vigil 2004).  Line widths observed in recombination lines are
typically 30 km s$^{-1}$ or greater.   Sensitive high resolution 18 cm (1.6
GHz) observations show weak emission extending to the NW-SE outside of the
ring, suggesting that, like the X-ray emission, some diffuse radio emission
has escaped the compact HII region.  Figure~\ref{2MASS} shows a bubble-like
feature associated with this emission in the NW.  The radio ring is not
well centered on IRS~2, likely indicating that the surrounding gas is not
homogeneous.

\section{Molecular Gas and Dust Continuum}

Detailed molecular line observations of RCW~38 are rare.  Gillespie et al.\
(1979) mapped a large region (10 $\times$ 5 pc) in CO 1-0 with a
3.2\arcmin\ beam.  They found two clouds separated by a bridge of emission.
Yamaguchi et al.\ (1999) also mapped CO 1-0 with similar resolution over a
larger region, confirming the earlier result, and estimated a mass of $1.5
\times 10^4$ M$_\odot$ for the two clouds as a whole.  The RCW~38 cluster
is directly associated with the eastern cloud.  Zinchenko et al.\ (1995)
mapped the eastern cloud over a 5 arcmin$^2$ region in CS 2-1 with
1\arcmin\ resolution.  Their maps show that CS, like the radio, follows a
ring-like structure, but close comparison indicates that the CS is located
exterior to the radio, tracing the dense molecular gas in which the compact
HII region is embedded.  The mass traced by CS is $\sim9000$ M$_\odot$
(correcting for their incorrectly assumed distance of only 0.7 kpc) with a
mean density of $10^4$ cm$^{-3}$.  A number of other high density tracers
have been observed with 1\arcmin\ or better resolution but maps are
lacking.

Deep absorption is observed in the main lines of the lowest rotational
levels of OH at 18 cm, and these have been used to measure the magnetic
field strength of 38$\pm$3 $\mu$G (Bourke et al.\ 2001).  High resolution
($\sim10\arcsec$) observations of OH show that the absorption is seen
against the radio ring, and is directly associated with the gas surrounding
the HII region, sharing the same velocity as the recombination lines
(T.Bourke, private communication).  No maser emission is observed in OH.
Water maser emission at 22 GHz is observed from a position near to IRS~1,
and it is time variable (Kaufmann et al.\ 1977; Batchelor et al.\ 1980;
Caswell et al.\ 1989).  Water maser emission is a sure sign of active star
formation, and high resolution observations are needed to precisely located
this emission.

The dust continuum has been mapped at wavelengths near to 1 mm (Cheung et
al.\ 1980; Vigil 2004).  Low resolution observations revealed only
one clump centered near IRS~1 (Cheung et al.\ 1980), with a peak density of
$6 \times 10^5$ cm$^{-3}$, a total mass of $3 \times 10^4$ M$_\odot$,  and
a visual extinction of 800 mag.  Higher resolution observations reveal the
ring-structure seen at other wavelengths (Vigil 2004), with a central hole
around IRS~2, but mostly peaking exterior to the radio ring.  In
particular, the position of peak intensity is west of the IRS~1 ridge and
associated with the extincted region seen in Figure~\ref{VLT}.  The mass
calculated by Vigil (2004) is an order of magnitude less than that of
Cheung et al.\ (1980), from similar fluxes.  As the morphology of the dust
emission observed by Vigil suggests that it is associated with the
radio ring, the extinction determined by Cheung et al.\ (1980) may
be an overestimate.

\section{Interpretation}

\begin{figure}[!ht]
\begin{center}
\includegraphics[draft=False,width=\textwidth, angle = 0]{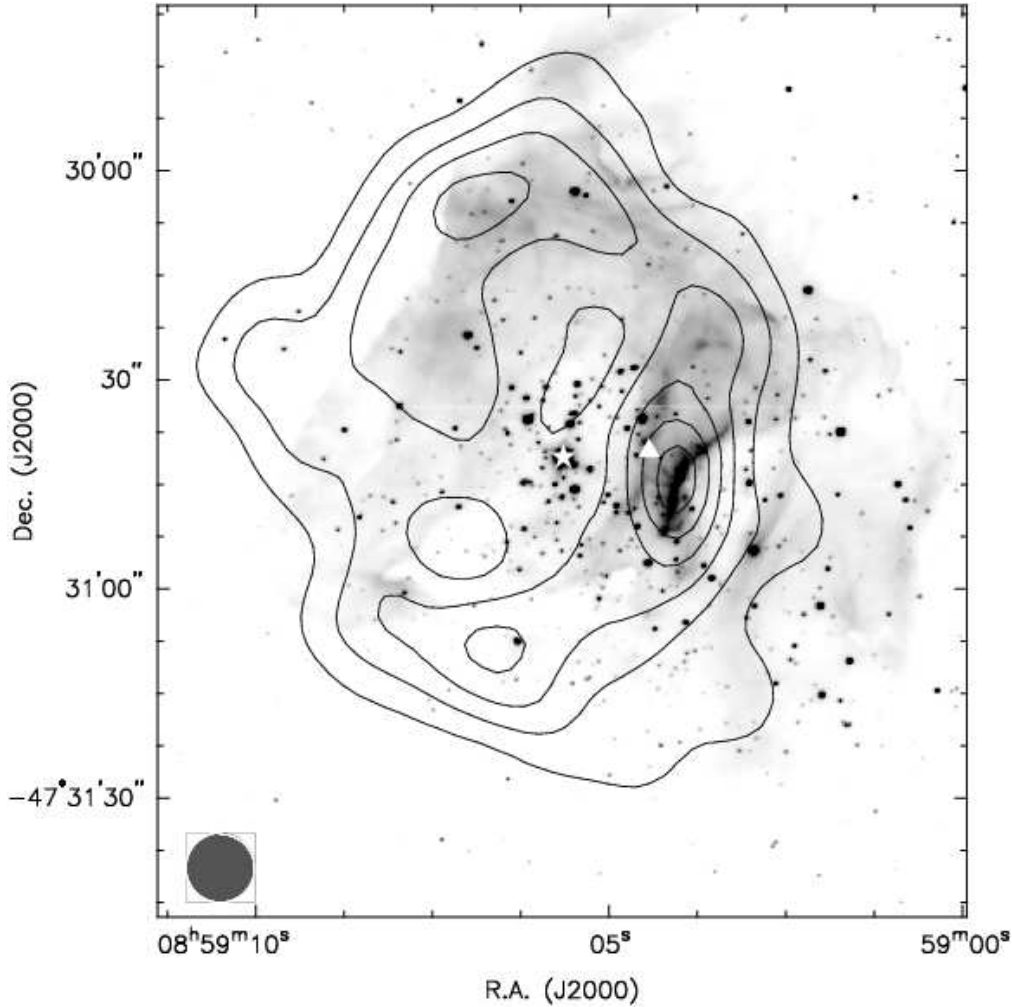}
\end{center}
\caption{4800 MHz (6 cm) radio continuum contours from the ATCA overlain on
the VLT K-band image shown in Figure~\ref{VLT} (30\arcsec\ = 0.25 pc).  The
white triangle indicates the location of the peak at 10~$\mu$m, IRS~1, and
the white star indicates the position of the O-stars IRS~2. The contour
levels are 15, 30, 45, 60, 75 and 90 percent of the peak emission of
3.5 Jy/beam.  The flux in the center region is low, between 15--30\% of the
peak flux level that is located near the IR ridge and IRS~1.  The synthesised
beam of 10\arcsec\ is shown at lower left. Adapted from Vigil 2004.}
\label{vigil3}
\end{figure}

The ring-like shape of the two dimensional radio and infrared continuum
image suggests the possibility of a shell--like structure surrounding IRS2.
The bulk of the emission seen in the cm images is concentrated in a ring
which would be the thick walls of a three dimensional shell.  The center
cavity would contain only the emission from the cover of the shell
perpendicular to the direction of observation (Figure~\ref{vigil3}).

We see two possible interpretations of the relationship among the
observational data -- one is wind driven (Smith et al.\ 1999), the other
supernova driven (Wolk et al.\ 2002).  The supernova scenario has, as its
key piece of evidence, a symmetrical ring of emission in the radio
continuum.  A second line of evidence for a supernova in this region is the
hot X-ray plasma with a power--law spectrum, which appears to be getting
harder as it moves away from the cluster center.  This is as one would
expect with synchrotron emission being accelerated from a compact object
near the cluster center.  The total X-ray luminosity is about $3 \times
10^{32}$~ergs~ sec$^{-1}$.  In these attributes it evokes the image of a
shell supernova remnant such as SN 1006. Further, if one assumes
equipartition, we derive a very reasonable magnetic field of about 4.9
$\mu$G and a reservoir of $2.2 \times 10^{44}$~ergs of accelerated
electrons, giving a lifetime of about 20,000 years.  But, the radio
continuum of RCW 38 reported here does not show the brightening at the
edges seen in SN 1006 (Reynolds \& Gilmore 1986).  The positive spectral
index of the radio continuum also favors thermal emission.  Further the
X-ray emission seen in shell supernovae tends to follow the form of the
radio emission (c.f. Allen et al.\ 2001).  Here, the plasma seems confined
by the continuum radio emission with the exception of the breakout to the
northwest.

This confinement naturally lends itself to a wind driven scenario.  Smith
et al.\ (1999) note that the central region around IRS~2 is dominated by
Ly$\alpha$ emission and is dust free.  This suggests that RCW 38 is a
compact HII region expanding into and confined by a non--homogeneous
molecular cloud.  The wind from the massive O-star (IRS~2) can give rise to
collisional shocks near to the star as described by Cant\'o et al.\ (2000).
Such shocks centered near IRS~2 would fill the cavity with a limb darkened
X-ray emission as seen.  But the derived thermal spectrum for the plasma is
much warmer than expected by these models ($\sim$10 keV ($\sim$ 100 MK) vs.
$\sim$ 1 keV ($\sim$ 10~MK)).  This high temperature may have to do with
confinement due to the surrounding molecular cloud.  Heated plasma will
tend to move outward and cool, but in the case of RCW 38 the molecular
cocoon prevents meaningful cooling and causes a ``greenhouse effect" on a
parsec scale. Smith et al.\ (1999) derived a similar result using
Ly$\alpha$ data.  Analysis of the smoothed X-ray data supports the
hypothesis that the molecular material is trapping the plasma that appears
to be partially heated by the trapped Ly$\alpha$ photons.  There appears to
be a break out of the plasma from the ring of the molecular gas in the
northwest corner while elsewhere the gas is confined much closer to the O
star.

Further evidence for this interpretation is found in other regions of
massive star formation.  Several massive star forming regions have been
seen to have much cooler diffuse X-ray emission.  Townsley et al.\ (2003)
report thermal spectra of below 1 keV for the Rosette Nebula and M17.
Other massive star forming regions, including the Arches cluster
(Yusef-Zadeh et al.\ 2002) and NGC~3603 (Moffat et al.\ 2002) have thermal
spectra above 3 keV.  While a commonality among these clusters is a massive
O star (O6 or earlier) the key difference between the former group and the
latter group seems to be the presence of absorbing (confining) material.
All the clusters with high temperature plasma have $N_H > 5 \times 10^{21}$
cm$^{-2}$ while the clusters with cooler plasma have less absorption.

The wind driven interpretation involves winds blowing from IRS~2 that are
excavating the material in its near vicinity creating this expanding shell,
seen as a radio continuum ring of emission.  IRS~2 is near the
center of the cavity and is likely the driving source.  However, IRS~2 is
displaced from the geometric center of the ring that is offset by about
18\arcsec\ to the northeast.  As noted earlier, this appears due to our
viewing angle at the system not being face on and the non-homogeneous
surroundings.  From inspection of the K--band image (Figure~5), an arc-like
feature, reminiscent of a bow shock, coincident with the bright radio
ridge, can be seen in the southwest of the ring.  It suggests either a
massive star forming region $\sim 10$\arcsec\ to the west, which is
creating a bow shock by deflecting the winds from IRS~2 around its edge, a
wind source from the southwest which could be counteracting the wind from
IRS~2 causing a compression ridge to build up along the edge of the ridge
-- IRS~1, or a dense globule being ablated (Bertoldi \& McKee 1997).
Indeed, there are about a half dozen X-ray sources within this arc, and
they all are found to have column densities in the range of 1.9 -
2.9$\times 10^{22}$~cm$^{-2}$, equivalent to about 10-15 A$_V$.
The southwest source, be it a massive star driving a wind or simply a dense
core, is preventing the expansion in the southwest direction, causing the
shell to expand non-uniformly and mostly toward the northeast, thus
explaining the offset of the center to the northeast.

From close inspection of the VLT K-band images, an arc-like feature
outlines the bright radio ridge (Figure~5).  This specific feature implies
dense gas that is likely photo-evaporating and could suggest that this is
a site of triggered star formation.  Additional evidence of active star
formation occurring just west of IRS~1 is a possible protostar found
embedded in the hydrogen gas and molecular jets eminating from this region
(DeRose et al.\ 2008), as well as the previously mentioned water masers.

The winds from IRS~2 appear to be confined by the surrounding molecular
cloud.  Assuming a typical O-star wind of 1000 km/s (Cant\'o et al.\ 2000)
and an age for IRS~2 of about 500,000 years suggests that a freely
expanding wind would have traveled about 500 pc since the star was formed.
The inner width of the shell is close to a tenth of a parsec, much smaller
than expected from a freely expanding wind.  Also, in the smooth X-ray
data, plasma is observed as far as 3.5 pc from IRS 2, far short of the 500
pc predicted for a freely expanding wind.

\section{Summary}

RCW~38 appears to be a blister compact HII region with winds originating
from the central O-star IRS~2.  These winds are excavating the mass in the
immediate vicinity of IRS~2 creating a shell--like structure, detected as a
radio continuum ring and also evident at infrared and millimeter
wavelengths.  A few hundred young low mass stars are found in the immediate
vicinty of IRS~2, and may be directly exposed to its winds and ionizing
radiation.  The region as a whole is estimated to contain around 2000 young
stars, with 30 OB star candidates.  The region of the ring to the west of
IRS~2, containing IRS~1, appears to be particularly active and is likely
the site of ongoing (triggered) star formation.  The IRS~1 ridge appears to
be the interface between the IRS~2 wind and either a similar wind from a
high mass star further to the west, or a dense clump of gas that is being
ablated by IRS~2.  Further observations are needed to test these scenarios.
In many ways, RCW~38 appears to be a younger more embedded version of the
ONC and deserves further study at higher angular resolution and across the
spectrum.

\acknowledgements 

We acknowledge many useful comments from the referee Leisa Townsley.
Jo\~ao Alves openly shared his unpublished data and ideas as part of our
ongoing collaboration on this region.  We thank Tom Megeath for fruitful
discussions.  We gratefully acknowledge the financial support from NASA
contract NAS8-39073 (CXC).  This work is based in part on observations made
with the Spitzer Space Telescope, operated by the Jet Propulsion
Laboratory, California Institute of Technology under a contract with NASA.
Support for this work was provided by NASA.



\begin{thebibliography}{}


\bibitem[Allen et al.(2001)]{2001ApJ...558..739A} Allen, G.~E., Petre, R.,
\& Gotthelf, E.~V.\ 2001, \apj, 558, 739

\bibitem[Aschenbach et al.(1999)]{1999A&A...350..997A} Aschenbach, B.,
Iyudin, A.~F., \& Sch{\"o}nfelder, V.\ 1999, \aap, 350, 997

\bibitem[Avedisova \& Palous(1989)]{1989BAICz..40...42A} Avedisova, V.~S.,
\& Palous, J.\ 1989, Bull. Astron. Inst. Czechoslovakia, 40, 42


\bibitem[Batchelor et al.(1980)]{1980AuJPh..33..139B} Batchelor, R.~A.,
Caswell, J.~L., Haynes, R.~F., Wellington, K.~J., Goss, W.~M.,
\& Knowles, S.~H.\ 1980, Australian Journal of Physics, 33, 139

\bibitem[Bertoldi \& McKee(1997)]{1997RMxAC...6..195B} Bertoldi, F., \&
McKee, C.~F.\ 1997, Rev. Mex. Astron. Astrofis. Conf. Series, 6, 195


\bibitem[Bourke et al.(2001)]{2001ApJ...554..916B} Bourke, T.~L., Myers,
P.~C., Robinson, G., \& Hyland, A.~R.\ 2001, \apj, 554, 916


\bibitem[Cant{\'o} et al.(2000)]{2000ApJ...536..896C} Cant{\'o}, J., Raga,
A.~C., \& Rodr{\'{\i}}guez, L.~F.\ 2000, \apj, 536, 896

\bibitem[Caswell \& Haynes(1987)]{1987AuJPh..40..215C} Caswell, J.~L., \&
Haynes, R.~F.\ 1987, Australian Journal of Physics, 40, 215

\bibitem[Caswell et al.(1989)]{1989AuJPh..42..331C} Caswell, J.~L.,
Batchelor, R.~A., Forster, J.~R., \& Wellington, K.~J.\ 1989, Australian
Journal of Physics, 42, 331

\bibitem[Cheung et al.(1980)]{1980ApJ...240...74C} Cheung, L.~H., Frogel,
J.~A., Hauser, M.~G., \& Gezari, D.~Y.\ 1980, \apj, 240, 74

\bibitem[Churchwell et al.(2004)]{2004ApJS..154..322C} Churchwell, E., et
al.\ 2004, \apjs, 154, 322

\bibitem[DeRose etal.(2008)]{2008AJ}  DeRose, K.L, Bourke, T.~L.,
Gutermuth, R.A., Wolk, S.~J., Megeath, S.T., Alves, J., N\"urnberger, D.\ 2008 \aj,
{\it submitted}

\bibitem[Epchtein \& Turon(1979)]{1979A&A....72L...4E} Epchtein, N., \&
Turon, P.\ 1979, \aap, 72, L4

\bibitem[Feigelson et al.(2007)]{2007prpl.conf..313F} Feigelson, E.,
Townsley, L., G{\"u}del, M., \& Stassun, K.\ 2007, {\em Protostars and Planets
V}, eds. B. Reipurth, D. Jewitt, \& K. Keil, 313

\bibitem[Frogel \& Persson(1974)]{1974ApJ...192..351F} Frogel, J.~A. \&
Persson, S.~E.\ 1974, \apj, 192, 351

\bibitem[Furniss et al.(1975)]{1975ApJ...202..400F} Furniss, I., Jennings,
R.~E., \& Moorwood, A.~F.~M.\ 1975, \apj, 202, 400

\bibitem[Gagn{\'e} et al.(2005)]{2005ApJ...628..986G} Gagn{\'e}, M.,
Oksala, M.E., Cohen, D.H., Tonnesen, S.K., ud-Doula, A. et al.\ 2005, \apj,
628, 986



\bibitem[Gillespie et al.(1979)]{1979MNRAS.186..383G} Gillespie, A.~R.,
White, G.~J., \& Watt, G.~D.\ 1979, \mnras, 186, 383

\bibitem[Gum(1955)]{1955MmRAS..67..155G} Gum, C.~S.\ 1955, \memras, 67, 155

\bibitem[Huchtmeier(1974)]{1974A&A....32..335H} Huchtmeier, W.\ 1974, \aap,
32, 335

\bibitem[Johnson(1973)]{1973PASP...85..586J} Johnson, H.~M.\ 1973, \pasp,
85, 586

\bibitem[Kaufmann et al.(1977)]{1977AJ.....82..577K} Kaufmann, P., Scalise,
E., Jr., Schaal, R.~E., Gammon, R.~H., \& Zisk, S.\ 1977, \aj, 82, 577



\bibitem[Lada \& Lada(2003)]{2003ARA&A..41...57L} Lada, C.~J. \& Lada,
E.~A.\ 2003, \araa, 41, 57

\bibitem[Ligori et al.(1994)]{1994MSAI.303.815L} Ligori S., Moneti A.,
Robberto M., Guarnieri M. D., Zinnecker H., 1994, Mem. Soc. Astron. Ital.,
303, 815


\bibitem[McGee \& Newton(1981)]{1981MNRAS.196..889M} McGee, R.~X., \&
Newton, L.~M.\ 1981, \mnras, 196, 889

\bibitem[Moffat \emph{et al.}(2002)]{2002ApJ...573..191M}Moffat, A.F.J.,
Corcoran, M.F., Stevens, I.R., Skalkowski, G., Marchenko, S.V. et al.\
2002, \apj, 573, 191.

\bibitem[Mizutani et al.(1987)]{1987MNRAS.228..721M} Mizutani, K., Suto,
H., Takami, H., Maihara, T., Sood, R.~K., Thomas, J.~A., Shibai, H., \&
Okuda, H.\ 1987, \mnras, 228, 721


\bibitem[Murphy(1985)]{1985PhDT........45M} Murphy, D.~C.\ 1985,
Ph.D.~Thesis, Massachusetts Institute of Technology

\bibitem[Muzzio(1979)]{1979AJ.....84..639M} Muzzio, J.~C.\ 1979, \aj, 84, 639

\bibitem[Muzzio \& Celotti de Frecha(1979)]{1979MNRAS.189..159M} Muzzio,
J.~C. \& Celotti de Frecha, M.~B.\ 1979, \mnras, 189, 159

\bibitem[Persson et al.(1976)]{1976ApJ...208..753P} Persson, S.~E., Frogel,
J.~A., \& Aaronson, M.\ 1976, \apj, 208, 753

\bibitem[Radhakrishnan et al.(1972)]{1972ApJS...24...49R} Radhakrishnan,
V., Goss, W.~M., Murray, J.~D., \& Brooks, J.~W.\ 1972, \apjs, 24, 49

\bibitem[Reynolds \& Gilmore(1986)]{1986AJ.....92.1138R} Reynolds, S.~P.,
\& Gilmore, D.~M.\ 1986, \aj, 92, 1138

\bibitem[Rho et al.(2004)]{2004ApJ...607..904R} Rho, J., Ram{\'{\i}}rez,
S.~V., Corcoran, M.~F., Hamaguchi, K., \& Lefloch, B.\ 2004, \apj, 607, 904

\bibitem[Rodgers et al.(1960)]{1960MNRAS.121..103R} Rodgers, A.~W.,
Campbell, C.~T., \& Whiteoak, J.~B.\ 1960, \mnras, 121, 103

\bibitem[Shaver \& Goss(1969)]{1969PASAu...1..280S} Shaver, P.~A. \& Goss,
W.~M.\ 1969, Proceedings Astron. Soc. Australia, 1, 280

\bibitem[Shaver \& Goss(1970)]{1970AuJPA..14...77S} Shaver, P.~A. \& Goss,
W.~M.\ 1970, Australian Journal of Physics, Astrophys. Suppl., 14, 77

\bibitem[Smith et al.(1999)]{1999MNRAS.303..367S} Smith, C.~H., Bourke,
T.L., Wright, C.M., Spoon, H.W.W.S., Aitken, D.K., et al.\ 1999, \mnras,
303, 367

\bibitem[Storey \& Bailey(1982)]{1982PASAu...4..429S} Storey, J.~W.~V. \&
Bailey, J.\ 1982, Proceedings Astron. Soc. Australia, 4, 429

\bibitem[Townsley et al.(2003)]{2003ApJ...593..874T} Townsley, L.~K.,
Feigelson, E.~D., Montmerle, T., Broos, P.~S., Chu, Y.-H., \& Garmire,
G.~P.\ 2003, \apj, 593, 874

\bibitem[Vigil (2004)]{Vigil04} Vigil M., 2004,  M.Sc. Thesis,
Massachusetts Institute of Technology



\bibitem[Whitney et al.(2004)]{2004ApJS..154..315W} Whitney, B.~A., et al.\
2004, \apjs, 154, 315

\bibitem[Wilson et al.(1970)]{1970A&A.....6..364W} Wilson, T.~L., Mezger,
P.~G., Gardner, F.~F., \& Milne, D.~K.\ 1970, \aap, 6, 364

\bibitem[Wolk et al.(2002)]{2002ApJ...580L.161W} Wolk, S.~J., Bourke,
T.~L., Smith, R.~K., Spitzbart, B., \& Alves, J.\ 2002, \apjl, 580, L161

\bibitem[Wolk et al.(2006)]{2006AJ....132.1100W} Wolk, S.~J., Spitzbart,
B.~D., Bourke, T.~L., \& Alves, J.\ 2006, \aj, 132, 1100

\bibitem[Yamaguchi et al.(1999)]{1999PASJ...51..791Y} Yamaguchi, R., Saito,
H., Mizuno, N., Mine, Y., Mizuno, A., Ogawa, H., \& Fukui, Y.\ 1999, \pasj,
51, 791

\bibitem[Yusef-Zadeh et al.(2002)]{2002ApJ...570..665Y} Yusef-Zadeh, F.,
Law, C., Wardle, M., Wang, Q.~D., Fruscione, A., Lang, C.~C., \& Cotera,
A.\ 2002, \apj, 570, 665

\bibitem[Zinchenko et al.(1995)]{1995A&AS..111...95Z} Zinchenko, I.,
Mattila, K., \& Toriseva, M.\ 1995, \aaps, 111, 95

\end{thebibliography}
\end{document}